\documentclass[manuscript]{aastex}

\newcommand{\FUSE}{{\em FUSE}}
\newcommand{\kms}{km s$^{-1}$}
\newcommand{\OVI}{\ion{O}{6}}
\newcommand{\CII}{\ion{C}{2}$^*$}
\newcommand{\CIII}{\ion{C}{3}}
\newcommand{\HI}{\ion{H}{1}}
\newcommand{\HeI}{\ion{He}{1}}

\newcommand{\lu}{photons cm$^{-2}$ s$^{-1}$ sr$^{-1}$}

\newcommand{\nh}{$N$(\HI)}
\newcommand{\rosat}{{\em ROSAT}}
\newcommand{\unit}[1]{\ifmmode {\rm\ #1} \else {$\rm #1$} \fi}
\newcommand{\percmsqr}{\unit{cm^{-2}}}
\newcommand{\persqrcm}{\unit{cm^{-2}}}

\shorttitle{Loop I/Local Bubble Interaction Region}
\shortauthors{Sallmen et al.}

\begin{document}


\title{\FUSE\ Observations of the Loop I/Local Bubble Interaction Region\altaffilmark{1}}
\altaffiltext{1}{Based on
observations made with the NASA-CNES-CSA Far Ultraviolet Spectroscopic Explorer.
FUSE is operated for NASA by the Johns Hopkins University under NASA contract
NAS5-32985.}



\author{Shauna M. Sallmen}
\affil{Department of Physics, University of Wisconsin - La Crosse, La Crosse, WI
54601}
\email{sallmen.shau@uwlax.edu}
\author{Eric J. Korpela}
\affil{Space Sciences Laboratory, University of California at Berkeley, Berkeley, CA, 94720}
\author{Hiroki Yamashita}
\affil{Department of Physics, McGill University, Montreal, QC, Canada, H3A 2T8}


\begin{abstract}
We used the \FUSE\ ({\it Far Ultraviolet Spectroscopic Explorer})
satellite to observe \OVI\ emission along two sightlines towards the
edge of the interaction zone (IZ) between the Loop I superbubble and
the Local Bubble.  One sightline was chosen because material in the
interaction zone blocks distant X-ray emission, and should thus do
the same for non-local \OVI\ emission. We measured an \OVI\ intensity
of $I_{\rm shadowed}$ = 2750 $\pm$ 550 \lu\ along this `Shadowed'
sightline, and $I_{\rm unshadowed}$ = 10800 $\pm$ 1200 \lu\ along the
other sightline.  Given these results, very little ($\lesssim$ 800 \lu)
of the emission arises from the near side of the interaction zone, which
likely has an \HI\ column density of about $4 \times 10^{20} \percmsqr$
along the `Shadowed' sightline.  The \OVI\ emission arising within Loop
I ($\sim$10$^4$ \lu) is probably associated with gas of $n_e \sim$ 0.1
cm$^{-3}$ and an emitting pathlength of $\sim$ 1.2 pc, suggesting it
arises at interfaces rather than from gas filling Loop I. In contrast,
the \CIII\ emission is similar along both sightlines, indicating that much
of the emission likely arises on the near side of the interaction zone.
\end{abstract}




\section{Introduction}

Gas in the galactic interstellar medium (ISM), including the galactic
halo, is heated by energy input from stellar winds and supernova
events. These processes are responsible for redistributing energy
and material throughout our galaxy, resulting in the formation of
new generations of stars. The non-uniform interstellar gas exhibits a
complex set of interacting shells, bubble-like structures, ``chimneys",
and worms that are seen as evidence of stellar energy input. Although
the physical state and evolution of these gas phases have been broadly
explained, it has not yet been determined whether the ISM is best
described by a three-phase model \citep{mckost77}, a galactic fountain
model \citep{shap76}, or a model with more isolated supernova remnants
\citep{coxsmith74,slavcox93}. There are still many outstanding problems
with these (and all other current) models of the ISM.

The far-UV (900-1200\AA) spectrum of diffuse interstellar emission
contains astrophysically important cooling lines: The \OVI\ doublet
($\lambda\lambda 1032,1038$) represents the dominant radiative cooling
mechanism for gas with temperatures between 10$^{5.4}$ and 10$^{5.7}$
K. The \CIII\ line ($\lambda 977$) is an important cooling mechanism in
gas with temperatures between 10$^{4.5}$ and 10$^{5.1}$ K \citep{chianti}.
Because of the high cooling rates due to these and other lines, gas
in this temperature range cools rapidly to lower temperatures, and
therefore we refer to gas in this temperature range as ``transition
temperature gas''.  Due to the rapid cooling, in order to be observed
this gas must be replenished, either from a source of higher temperature
gas cooling through this temperature range, shock heating of cooler gas,
conductive heating in a boundary between hot and cold gas, or turbulent
mixing of hot and cold gas \citep{mckost77, spitzer90, slavin93}.

The \OVI\ ion, characteristic of gas with a temperature of
$\sim$300,000~K, is a sensitive probe of transition temperature gas in our
galaxy.  In recent years, several detections of galactic \OVI\ emission
have been made with the \FUSE\ ({\it Far Ultraviolet Spectroscopic
Explorer}) satellite.  Typical values at 1032 \AA\ in directions with
low \nh\ are 2000-3300 \lu\ (LU) \citep{dix01, shel01, shel02, shel03,
wel02}, although a recent measurement of halo gas was somewhat higher
\citep{shel07}.  The \OVI\ survey of \citet{dix06} sampled 183 sightlines,
29 at 3-$\sigma$ significance, with a median of 3300 LU. The median value
of all 3-$\sigma$ upper limits is 2600 LU.  Until recently, the galactic
location of this hot-gas emission was unknown.  \citet{shel03} concludes
that the local ($<230$ pc) contribution to this emission is negligible,
with a 2-$\sigma$ upper limit of 500 LU in the direction ($l$,$b$) =
(278.6\arcdeg, -45.3\arcdeg) (later revised to 600 LU by \citet{dix06}).
It therefore appears likely that most emission at high galactic latitudes
arises from hot gas in the galactic halo.

Superbubbles are extremely large structures in the ISM, believed to be
blown by the combined energy output of a cluster of stars. Such regions
provide an important diagnostic of the processes by which supernovae and
stellar winds control the overall evolution of our galaxy. Superbubbles
are expected to be filled with hot emitting gas.  \OVI\ emission
intensities towards supernova remnants (SNRs) and superbubbles can
be significantly higher than in the general ISM. \citet{dix06} noted
two sightlines in their survey that fit this category, each with OVI
intensities exceeding 8000 LU. {\it SPEAR} detected the extremely high
value of 180,000 LU towards the Vela SNR \citep{nish06}, and nearly 7000
LU towards the edge of the Orion-Eridanus Superbubble \citep{kreg06}.

In this paper, a shadowing strategy is used to determine the location
of hot emitting gas towards the Loop I superbubble. Observations for the
adjacent directions were made to compare the intensity of emission from
each sightline. Since one sightline contains material which significantly
blocks the distant emission and the other does not, intensities from
each sightline are different. This difference in intensity tells us the
general location of hot emitting gas.

In Subsection~\ref{subsect:region}, we outline the
important characteristics of the region of our observations.
Sections~\ref{sect:observations} and \ref{sect:results} describe the
observations and results.  In Section \ref{sect:discussion} we discuss
what our observations tell us about physical conditions in the Loop I
superbubble and the Local Bubble.

\subsection{Description of Region}\label{subsect:region}

Loop I is a large-scale structure first discovered in the radio continuum
sky \citep{berk71}.  The 116\arcdeg $\pm$ 4\arcdeg\ radio ring is centered
on ($l$,$b$) = (329\arcdeg $\pm$ 1.5\arcdeg, 17.5\arcdeg $\pm$ 3\arcdeg).
It is widely believed to be a superbubble blown by strong stellar winds
and supernovae of the Sco-Cen OB association ($\sim$170 pc away).  X-ray,
neutral hydrogen, and optical absorption measurements are consistent
with a shell of radius $\sim$ 100 pc centered $\sim$ 130 pc away in the
direction (l,b) = (330\arcdeg, 15\arcdeg), with the receding shell $\la$
212 pc away \citep{nish99}.  Note, however, that \citet{wel05} used {\it
HST-STIS} UV absorption spectra towards the approximate center of Loop I,
($l$,$b$) = (330\arcdeg, 18\arcdeg), and estimated that the approaching
and receding walls of Loop I are about 90 pc and 150-180 pc, respectively.

\citet{ea95} identified an annular structure (see Figure 3 of their
paper) seen inside the Loop I neutral-hydrogen shell in the \HI\ map
of \citet{dl90}. This feature is interpreted as the interaction zone
(IZ) between Loop I and our Local Bubble.  Using data compiled by
\citet{frus94}, \citet{ea95} determined that its distance is $\sim$
70 pc and $N_H$ jumps from less than $10^{20} \persqrcm$ to over $7
\times 10^{20} \persqrcm$ at this distance.  If Loop I is spherical,
this distance is inconsistent with the aforementioned results of
\citet{wel05}.  However, \citet{cor03} used color excesses to indicate
that the interaction zone is twisted and folded, with a transition to
higher reddening values occurring at distances ranging from 60 pc to 180
pc, depending on direction, a result further supported by absorption-line
studies of the region interior to the annulus \citep{cor04}.

\begin{figure}
\plotone{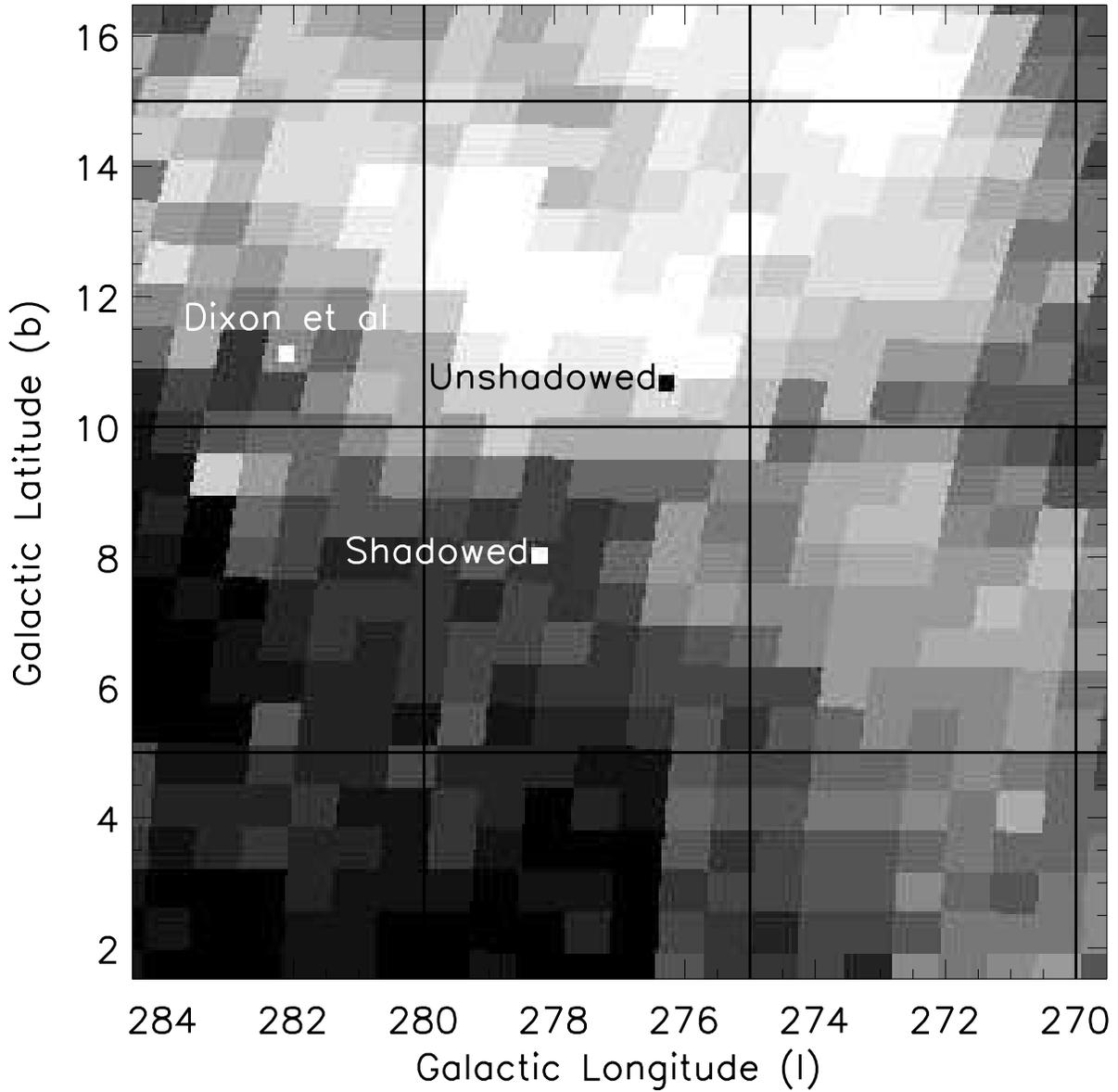}
\caption{1/4 kev \rosat\ map \citep{snowden97} showing our sightlines, as well
as the location of an \OVI\ detection by \citet{dix06}. 
More X-rays are received from brighter areas.
Note the X-ray shadowing due to the neutral hydrogen in the IZ.
\label{fig:ROSAT_small}}
\end{figure}

Our observations lie along the edge of the IZ near ($l$,$b$) =
(277\arcdeg, +9\arcdeg).  The $\frac{1}{4}$ keV X-ray map for
a small region surrounding our observations is shown in Figure
\ref{fig:ROSAT_small} \citep{snowden97}. The `Shadowed' sightline
intersects the neutral-hydrogen interaction zone, while the `Unshadowed'
sightline passes through an adjacent region of low neutral-hydrogen
column density. The neutral gas blocks high-energy photons, causing the
X-ray shadowing effect seen in the \rosat\ image. This same material will
block distant \OVI\ emission. Comparison of the two sightlines gives us
the opportunity to distinguish between local and distant emission.

\begin{figure}
\plotone{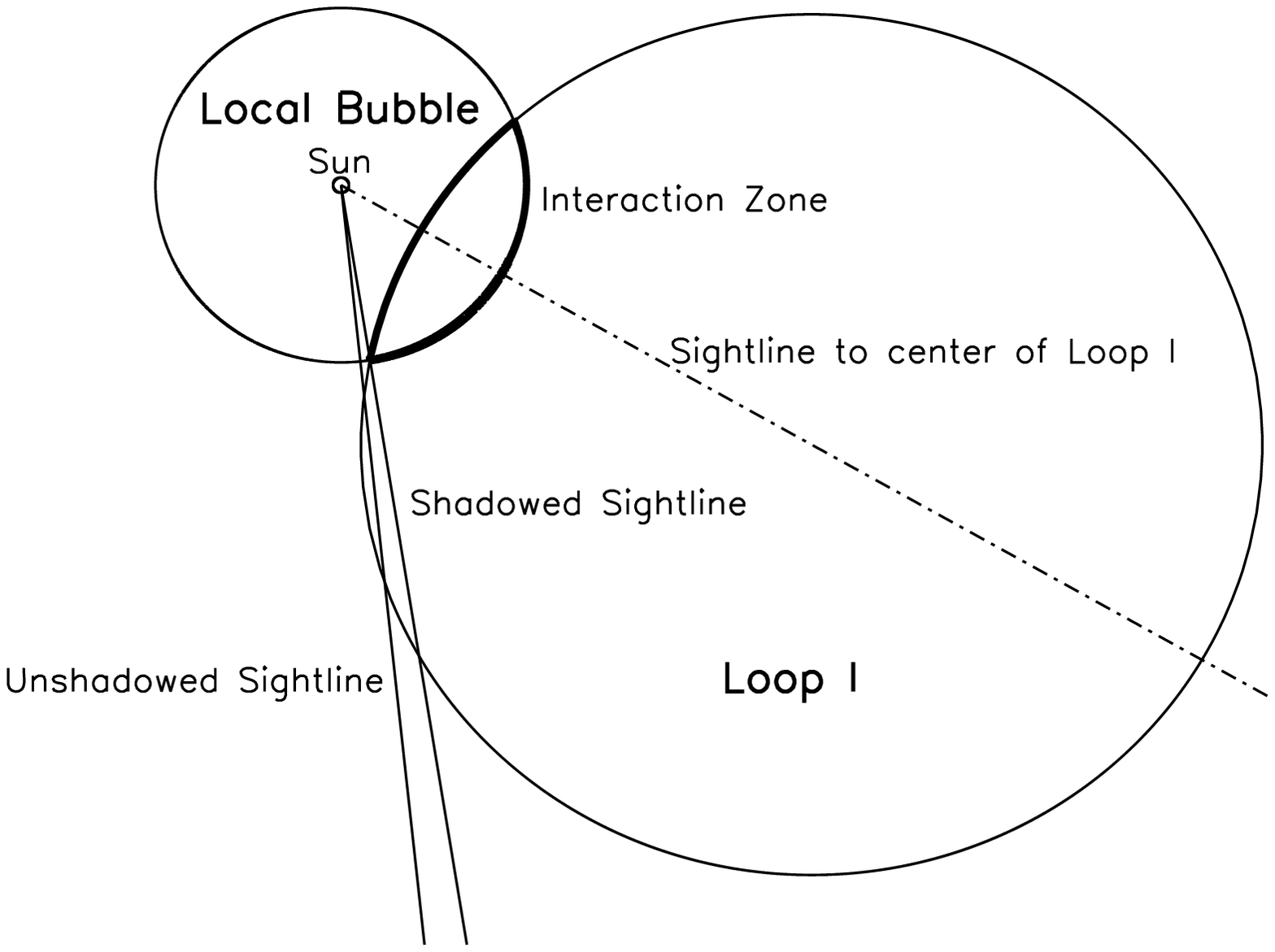}
\caption{Schematic Diagram showing the relative locations of the Sun, Local
Bubble, Loop I Superbubble, interaction zone, and our sightlines. 
\label{fig:schematic}}
\end{figure}

Figure \ref{fig:schematic} schematically shows the Local Bubble,
Loop I superbubble, and the interaction zone between them. The
lines of sight are shown for each direction of observation. \\

\section{Observations}\label{sect:observations}


\FUSE\ is composed of four separate optical systems. Two employ LiF
optical coatings and are sensitive to wavelengths from 990 to 1187
\AA, while the other two use SiC coatings, which provide reflectivity
to wavelengths as short as 905 \AA.  The four channels overlap in
the astrophysically important 990-1070 \AA\ region.  For a complete
description of \FUSE, its mission, and its in-flight performance, see
\citet{Moos:00} and \citet{Sahnow:00}.

The \FUSE\ spectrum of the `Shadowed' sightline was obtained in 3
observations (C1640401, C1640402, C1640403).  The first two were obtained
on 2002 June 27-29, and the third on 2003 Februray 18/19.  Each exposure
was centered on ($l$,$b$) = (278.23\arcdeg, +8.02\arcdeg). The total
usable exposure time for detector 1 was (after screening by the pipeline)
about 60 ksec, with 40.6 ksec obtained in orbital night. Detector 2 had
an additional 24.5 ksec of usable data, of which 4.4 ksec were obtained
in orbital night.  Data for the `Unshadowed' sightline (C1640301),
centered on ($l$,$b$) = (276.26\arcdeg, +10.692\arcdeg), were obtained
on 2004 April 3/4. The total usable exposure time for both detectors
was about 43 ksec, with 30 ksec obtained during orbital night.

Data were reduced using CalFUSE v3.1.3.  The data reduction process
includes burst screening, removal of data obtained during passages through
the South Atlantic Anomaly or at low earth-limb angles, pulse height
screening, corrections for spacecraft motions, dead-time corrections,
and spectral binning to 0.013 \AA\ \citep{dix07}.  The pipeline also
performs an initial wavelength calibration, as well as flux calibration.
Background removal was suppressed.  After testing, default pulse height
ranges were deemed appropriate and used for all detector segments.

The exposures for each observation were added together prior to final
spectral extraction.  In addition, all data were processed once including
photons from both orbital day and orbital night, and once including
only those from orbital night.  For SiC channels, scattered sunlight
contaminates the daytime spectra, so only orbital night spectra were used.

For each detector segment / optical channel, we determined the zero point
of the wavelength scale by measuring the observed heliocentric
wavelengths of airglow lines for each observation, and applying the
measured offset (no other term was deemed necessary) to the spectrum.
After this, multiple observations were combined on this corrected
heliocentric scale, if necessary.  This should correct for possible velocity
shifts between the various segments. Remaining systematic errors in
wavelength are estimated to be $\sim$15 \kms\ for LiF 1a, LiF 2a, and SiC 2a 
(corresponding to 0.052 \AA\ for LiF 1a). We were unable to accurately
wavelength calibrate the SiC 1a spectrum.

\section{Spectral Analysis and Results}\label{sect:results}

\begin{figure}
\plotone{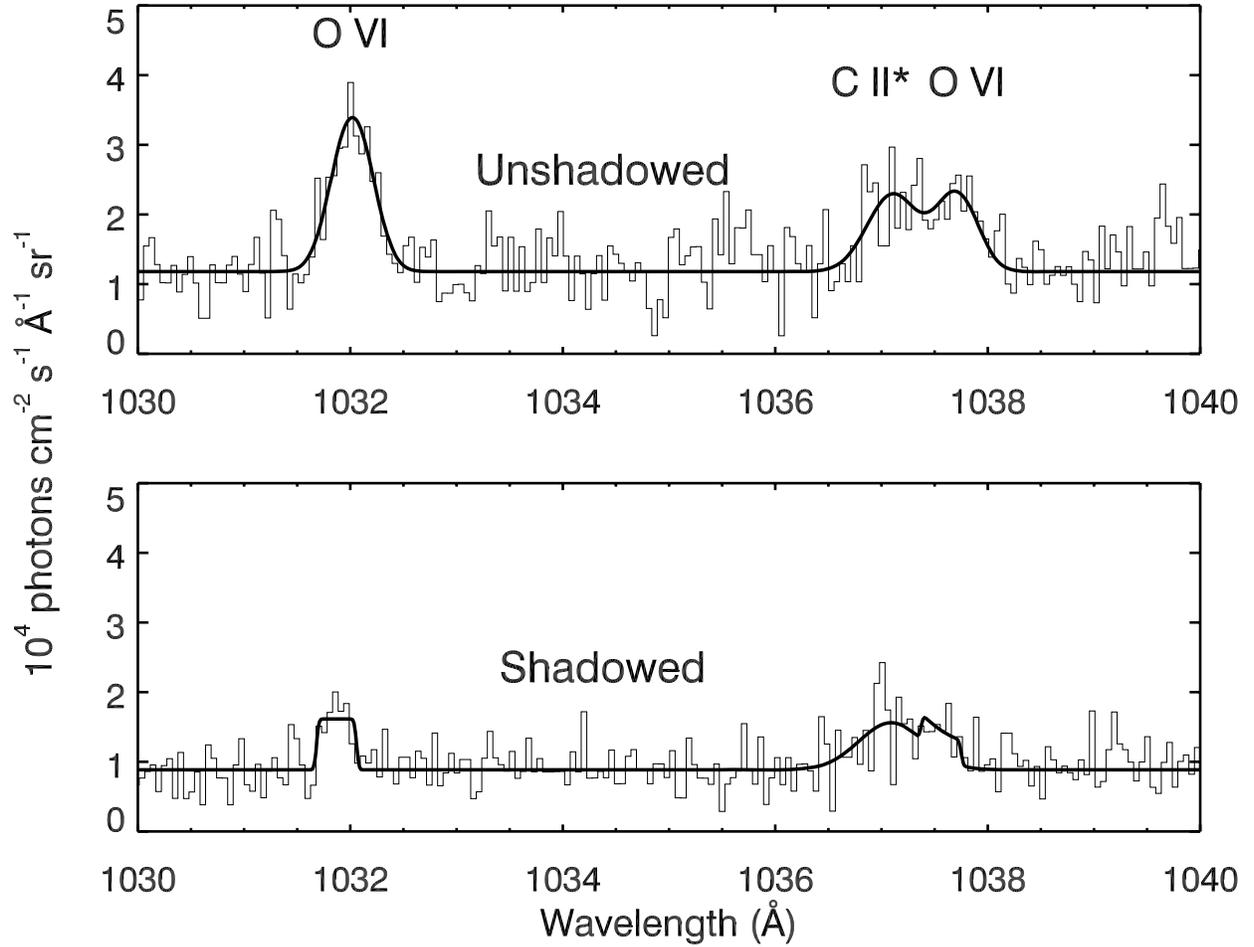}
\caption{Night-only \OVI\ spectrum for the two sightlines. Spectral data have been
binned to 0.052 \AA. The dark solid line represents our
fitted model, as described in the text.\label{fig:OVIspec}}
\end{figure}

The resulting spectra for the two sightlines are shown for the 1030-1040\AA\
wavelength region in Figure \ref
{fig:OVIspec}. Only nighttime spectra are shown, as several others have
noted an apparent airglow feature (possibly the second-order diffraction
peak of \HeI\ at 515.62\AA) near 1032 \AA\ \citep{wel02,shel07}. No such
feature is detectable in our data, and the nighttime measurements are
entirely consistent with spectra produced by including daytime data. The
two emission lines of the \OVI\ doublet are clearly visible. The 1038 \AA\
line is blended with a 1037 \AA\ \CII\ emission line. The intensities
of these lines were measured as follows.

Each emission line was assumed to be described by a 106 \kms\ tophat
(image of LWRS aperture) convolved with a Gaussian characterized by
$\sigma_G$, which includes both intrinsic and instrumental contributions
\citep{dix06}.  The 106 km/s appears to be nearly correct for LiF1a,
but the airglow lines appear somewhat narrower for other detector
segments.  The data were binned by 4 pixels prior to fitting, resulting
in a pixel scale corresponding to 15 km/s. The errors were also smoothed
so as to avoid 0 values, without removing significant structure.
The lines were fit using the IDL-based MPFIT routines developed by Craig
Markwardt\footnote{http://cow.physics.wisc.edu/$\sim$craigm/idl/idl.html}.
Initially, only the 1032 \AA\ line was included in the fit.  Since the
\CII\ emission line at 1037 \AA\ is mixed with \OVI\ emission at 1038 \AA,
we assume that the 1038 \AA\ \OVI\ emission line is half the intensity
of the \OVI\ emission line at 1032 \AA.  This assumption is true for
optically thin gas, and yields reasonable results in this case. This
\OVI\ fitted model was removed from the entire spectrum, leaving the
\CII\ emission line. The \CII\ emission line was then fitted in the
same way as the \OVI\ emission line. The reasonableness of the \CII\
fits validates this procedure. The complete model is shown over the
original spectrum in the figure.

The results of the \OVI\ fits are reported in Table \ref{table:OVI}. The
results of the \CII\ fits are included in Table \ref{table:otherlines}.
The heliocentric wavelengths yield a heliocentric radial velocity,
and have been converted to the standard Local Standard of Rest (LSR) in
Table \ref{table:OVI}.  The 1-$\sigma$ random error bars for each model
parameter have been determined using the error estimation prescription
of \citet{Press:88}, as described in \citet{dix01}.  Note that the
flux calibration uncertainty of $\sim$ 10\% \citep{Sahnow:00}, when
combined with uncertainty in the solid angle of the LWRS aperture,
results in a systematic uncertainty of $\sim 14$\% \citep{shel01}.
The width of the convolving Gaussian (reported as $\sigma_G$ in Table
\ref{table:OVI}) includes both the intrinsic width of the emission
line, and an instrumental contribution of 25 km/s \citep{dix06}. The
FWHM reported in Table \ref{table:OVI} have been adjusted for this
instrumental effect, without associating any error with the correction.

As noted earlier, we see no sign of contamination by airglow in the
full dataset.  However, in what follows we restrict ourselves to the
results from the night-only spectra. In general, measurements from the
full dataset are similar to those from the night-only spectra.

We also searched all the night-only spectra for any other detections at
the $\sim3\sigma$ level. For every wavelength region, we considered the
detector segment with the highest effective area. We assumed that the
instrumental response was a tophat with width 0.365\AA\ (corresponding
to 106 \kms\ at 1032 \AA).  We took a running average of each spectrum
using this width, adjusting the errors appropriately. After removing a
local median estimate of the continuum / background for each point, we
statistically determined the 99.7\% confidence limit of the data.  Since
the errors are likely underestimated due to the low number of counts,
this does not correspond precisely to a 3$\sigma$ level.  Any spectral
data points exceeding this level (often airglow) were investigated, and
the results for all spectra/detectors at that wavelength were compared.
Line-fitting and error determination was done (as described earlier)
for both sightlines if warranted for either. In some cases we excluded
lines which appeared to be marginal detections in low sensitivity
spectra, but did not appear in higher sensitivity spectra taken with
other detector segments.  The results of the fits are reported in Table
\ref{table:otherlines}, including the LSR radial velocity for features
detected on both sightlines.

Note that we have searched for features that result from a filled (or
nearly filled) aperture, but minimal intrinsic width. We might therefore
have missed significant broad features in the spectra.  As an example,
consider the \CIII\ line, which met our criteria in the `Shadowed'
sightline but not the `Unshadowed' one, where it is broader (see Table
\ref{table:otherlines}).

To determine upper limits on astrophysically interesting emission lines,
we first estimated the 95\% confidence limit for each point, evaluated
in the same fashion as discussed above. We then determined the median
value of this limit in a region 0.5 \AA\ wide. These values are quoted
in Table \ref{table:otherlines}.  Any feature for which some points lay
above the 95\% confidence level in this region were fit. The results
of the fits are included in the table, however the widths were poorly
determined and are not tabulated.  From these measurements, we tabulate
the radial LSR velocity only for intensity measurements that are unlikely
to be contaminated with airglow.




\section{Discussion}\label{sect:discussion}

\subsection{\OVI\ Intensities within Loop I}

The observed intensities for the `Shadowed' and `Unshadowed' sightlines
are $I_{\rm shadowed}$ = 2750 $\pm$ 550 \lu\ and $I_{\rm unshadowed}$
= 10800 $\pm$ 1200 \lu\ (LU).  To interpret our data, we made several
assumptions and used the following two equations for the intensity of
emission from each sightline:

\[
I_{\rm shadowed} = I_{\rm LB} + I_{\rm beyond} e^{-\tau_s} 
\]
\[
I_{\rm unshadowed} = I_{\rm LB} + I_{\rm beyond} e^{-\tau_{us}}
\]
where $I_{\rm LB}$ indicates the intensity of emission from the Local
Bubble (more specifically, from the near side of the IZ) and $I_{\rm
beyond}$ indicates the intensity of emission from beyond any intervening
material along the `Unshadowed' sightline. This material has optical
depth $\tau_s$ along the `Shadowed' sightline and $\tau_{us}$ along the
`Unshadowed' sightline.  Note that $\tau_{us}$ may be non-zero, because
that sightline passes through the boundaries of both the Local Bubble
and Loop I (see Figure \ref{fig:schematic}).  Most significantly, we
are assuming that $I_{\rm LB}$ and $I_{\rm beyond}$ are the same along
both the `Shadowed' and `Unshadowed' sightlines, and that the difference
between the optical depths along the two sightlines ($\tau_{\rm IZ}$)
is due entirely to material in the IZ.  It is unclear how patchy the
\OVI\ emission or absorbing material might be.  The possible limitations
of these simplifying assumptions must be kept in mind throughout the
discussion below.

Because we expect the IZ to contain most of the absorbing material, for
simplicity in this section we assume that $\tau_{us}$ = 0 and $\tau_s
= \tau_{\rm IZ} = \tau$. We will relax this assumption in Section
\ref{subsect:physical}.  To get a lower limit on the \OVI\ emission
interior to Loop I, we assumed (Case 1) that the interaction zone (IZ)
was completely opaque ($\tau = \infty$). In this case, $I_{\rm LB}$ =
2750 $\pm$ 550 LU and $I_{\rm beyond}$ = 8050 $\pm$ 1320 LU.  In Case 2,
we assumed that no emission arose within the Local Bubble ($I_{\rm LB}
= 0$ LU), reflecting the results of \cite{shel03}. This yields $I_{\rm
beyond}$ = 10800 $\pm$ 1200 LU, optical depth $\tau$ = 1.4 $\pm 0.2$ and
\nh$_{\rm IZ} = 4.1 \pm 0.7 \times 10^{20} \percmsqr$ (assuming \nh\ = 3
$\times$ 10$^{20} \persqrcm$ for $\tau$ = 1; \citet{sasseen02}).  Finally,
for Case 3 we assumed that the shadowing material in this region had the
average properties of the interaction zone as determined by \citet{ea95}.
Under these assumptions \nh$_{\rm IZ} = 7 \times 10^{20} \persqrcm$,
$\tau = 2.3$, $I_{\rm LB}$ = 1890 $\pm$ 620 LU and $I_{\rm beyond}$ =
8910 $\pm$ 1460 LU.

The extreme Case 1 is unlikely to be correct, especially in light
of the \citet{shel03} upper limit to Local Bubble emission of 600 LU
(revised by \citet{dix06}), but the derived lower limit does indicate that
substantial emission must be arising from beyond the IZ.  Case 2 provides
a lower limit on the IZ \HI\ column density that is consistent with our
results and assumptions, since a smaller value for \nh\ yields negative
emission within the Local Bubble.  Note that assuming the Local Bubble
contribution is $I_{\rm LB} = 600$ LU does not significantly change the
calculated value of $\tau$ or \nh$_{\rm IZ}$ for this Case.  The Local
Bubble contribution resulting from Case 3 is significantly larger than the
\citet{shel03} upper limit. Either the actual \HI\ column density of the
IZ along our sightline is less than the value reported by \citet{ea95},
the Local Bubble \OVI\ emission is substantially non-uniform, or there
is more \OVI\ emission arising in interfaces on the near side of the IZ
than elsewhere in the Local Bubble. In the latter case, the interaction
between Loop I and the Local Bubble may have compressed or otherwise
altered Local Bubble material in this region. As we will discuss below
and in Section \ref{subsect:physical}, the Case 3 \HI\ column density
is also inconsistent with other estimates of \nh$_{\rm IZ}$.

Recent data from Reis \& Corradi (2007, personal communication)
\nocite{reis07} suggest that the interaction zone along this line of
sight has an E($b-y$) extinction of 0.03, which would be equivalent to a
hydrogen column of $\sim 2\times10^{20} \persqrcm$ \citep{knude78}. This
is a factor of two below the lower limit to \HI\ column density consistent
with our simple model (see Case 2 above), and appears inconsistent with
our results and assumptions.  However, since our particular line of
sight was chosen because of a notable X-ray shadow, it is likely that it
represents a more dense portion of the interaction zone than is typical,
and so has a higher column density than is typical.

We also considered the \cite{dl90} measurements of total (to infinite
distance) \HI\ column density along the line of sight for both
sightlines.  The nearest Dickey \& Lockman (hereafter DL) measurements
give \nh$_{unsh}$ = $1.259 \times 10^{21} \persqrcm$ and \nh$_{shad}$ =
$1.412 \times 10^{21} \persqrcm$.  The difference between them is $1.53
\times 10^{20} \persqrcm$, which we use to estimate $\tau$, assuming that
this difference arises solely within the IZ. Then our simple model yields
an unphysical negative value for the Local Bubble intensity, suggesting
that the total neutral hydrogen differences in the two directions are
not due only to the interaction zone, but also include differences in
material beyond the IZ; we cannot necessarily use these measurements
to estimate $\tau_{\rm IZ}$. Interpolating over the four nearest DL
measurements\footnote{Retrieved using {\it CHANDRA}'s Colden Neutral
Hydrogen Density Calculator at http://cxc.harvard.edu/toolkit/colden.jsp}
gives \nh$_{unsh}$ = $1.018 \times 10^{21} \persqrcm$ and \nh$_{shad}$ =
$1.416 \times 10^{21} \persqrcm$. The difference between these is $3.97
\times 10^{20} \persqrcm$, consistent with Case 2 above.  However,
because DL is a composite of multiple surveys averaged over 1 degree
bins and because galactic column densities can vary significantly over
arcminute scales, there are potentially large errors when comparing to
emission in two 30\arcsec\ fields.

\citet{dix06} detected \OVI\ emission with an intensity of 4000
$\pm$ 1300 LU along a nearby sightline at ($l$,$b$) = (282.1\arcdeg,
+11.1\arcdeg).  As can be seen in Figure \ref{fig:ROSAT_small}, this
sightline also lies near the edge of the IZ, and has less extinction
at $\frac{1}{4}$ keV than our `Shadowed' sightline, but more than our
`Unshadowed' sightline.  Not surprisingly, the observed \OVI\ emission
lies between our two measurements. However, the 1-$\sigma$ errors are
such that this sightline could, in principle, have \OVI\ emission less
than that along our `Shadowed' sightline. As a result, this sightline
is of limited use in further constraining our results.

In all cases considered, the local contribution to the emission is small,
indicating that most of the emission lies beyond our Local Bubble.
In Case 2 (which we prefer), the measured \OVI\ emission along the
`Shadowed' sightline largely originates beyond the IZ. The estimated
intensity of emission from beyond the IZ is significantly larger than
the typical measurement ($\sim$ 2500 LU) for directions of low hydrogen
column density. Given the total \nh\ $\sim$ 10$^{21} \percmsqr$ along our
sightlines, \OVI\ emission from the thick-disk or galactic halo would be
attenuated by a factor of $\sim 20$ ($e^{-3} \sim 0.05$).  To provide
a significant portion of the observed emission, such a component would
require intrinsic intensities of 10$^5$ \lu, which could only arise from
highly over-pressure regions such as young SNR, as seen in \OVI\ maps
generated by {\it SPEAR} (\cite{edel07}, \cite{korpela06}).  Since no
known young SNR exist on this line of sight, we conclude a significant
portion of our detected emission must be associated with the Loop I
Superbubble itself.  For the remainder of the paper, we assume that all
detected \OVI\ emission arising beyond the IZ originates within Loop I.

\subsection{Physical Conditions within Loop I}{\label{subsect:physical}}

It is possible to use the \rosat\ $\frac{1}{4}$ keV (R12) and
$\frac{3}{4}$ keV (R34) background measurements to estimate the difference
in hydrogen column due to the IZ along each line of sight. Using a
0.1\arcdeg\ radius circle centered on our sightlines, the \rosat\ count
rates (10$^{-6}$ counts s$^{-1}$ arcmin$^{-2}$) along the `Shadowed'
sightline are R12 = 749 $\pm$ 175 and R34 = 227 $\pm$ 86 while those for
the `Unshadowed' sightline are R12 = 1423 $\pm$ 165 and R34 = 180 $\pm$
61 \citep{snowden97}\footnote{Retrieved using the X-Ray Background
Tool: http://heasarc.gsfc.nasa.gov/cgi-bin/Tools/xraybg/xraybg.pl}.
We modeled the region as a {\it CHIANTI} equilibrium plasma model
\citep{chianti0,chianti}.  The output spectrum of this model is absorbed
along the `Unshadowed' line of sight by a non-IZ \HI\ column, and on
the other by a combination of the non-IZ \HI\ column and an additional
column density due to the IZ.  As noted earlier, the boundaries of the
Local Bubble and Loop I may well provide material that attenuates emission
arising from within Loop I, even along the `Unshadowed' sightline. By 
assuming that both the X-ray plasma temperature and emission measure are
the same on both lines of sight, we use the X-ray band intensities to
derive a plasma temperature of $(1.26\pm0.06)\times 10^{6}$ K, a non-IZ
column density of $(1.5\pm0.2)\times10^{20}$ \percmsqr, and an IZ column
density of $(4.0\pm0.2)\times 10^{20}$ \percmsqr. (In this circumstance the
errors represent the $\Delta\chi^2=1$ limits of the absorbed plasma
models, which includes the statistical errors of the X-ray count rates.)
The estimated IZ column density is consistent with the value we obtained
for Case 2 by assuming none of our observed \OVI\ emission arises within
the Local Bubble, although that result assumed there was no extinction
along the `Unshadowed' sightline.  The hot X-ray emitting plasma in
this model does not produce significant OVI emission ($I_{OVI}\lesssim$
10 LU) and is, therefore, not sufficient to explain the observed \OVI\
emission without significant additional plasma at lower temperature.

We now relax our earlier assumption that $\tau_{us} = 0$, and instead use
the values corresponding to the output of the CHIANTI models: $\tau_{us} =
0.50 \pm 0.07$, $\tau_{\rm IZ} = 1.33 \pm 0.07$, and $\tau_s = \tau_{us}
+ \tau_{\rm IZ} = 1.83 \pm 0.09$.  This results in $I_{beyond} = 18000
\pm 3300$ LU, and $I_{LB} = -130 \pm 900$ LU. The negative Local Bubble
emission is unphysical, but the result is entirely consistent with no
emission from this side of the IZ. Based on this model, $\lesssim$ 800 LU
(1-$\sigma$ upper limit) of \OVI\ emission arises on the near side of the IZ. 
Compared with the earlier estimate
from Case 2, allowing for extinction by intervening material increases
the amount of \OVI\ emission arising within Loop I by $\sim$60\%.

In order to determine how much of the \OVI\ emission arises from hot gas
filling the Loop I superbubble, and how much from hot gas mixing with
cooler gas at the interfaces within the interaction zone, we estimate
the pathlength of the \OVI-emitting gas as follows.  We first assume that
the \OVI-emitting plasma is near the temperature of peak \OVI\ emission
($\sim 3\times 10^5$ K) and convert the Loop I \OVI\ emission to an
emission measure.  Using 18000$\pm$3400 LU for the extinction corrected
Loop I \OVI\ intensity results in an emission measure of $0.012\pm0.002$
cm$^{-6}$ pc.  Temperatures away from the \OVI\ peak require higher
emission measure, with $2.4\times10^{5} K$ and $3.2\times10^{5} K$
representing the points at which the required emission measure would
double.

\citet{dix06} found two types of \OVI-emitting gas within the galaxy:
gas with densities of $n_e \sim$ 0.01 cm$^{-3}$ and gas with $n_e \sim$
0.1 cm$^{-3}$. These two densities result in \OVI-emitting pathlengths
of $\sim$120 pc and $\sim$1.2 pc, respectively.  If we further assume
pressure equilibrium between the X-ray emitting plasma and the lower
temperature \OVI-emitting plasma, as would be expected if the \OVI\
originates in interfaces between hot and cool gas, we can derive the
relative filling factors of the X-ray and \OVI\ plasmas within Loop I. We
performed this calculation using the appropriate X-ray temperature,
X-ray and \OVI\ emission measures as found above.  We find that given
the above values and their error, the Loop I \OVI-emitting path is
between $0.71$ and $1.2$ times the X-ray emitting path if the
\OVI-emitting plasma is near the \OVI\ peak.  If the \OVI-emitting plasma
is not at the \OVI\ peak, the required path length could be substantially
larger.  Similarly, if the pressure in the X-ray emitting gas is higher
than that of the \OVI-emitting gas, the relative \OVI-emitting path
is correspondingly larger by a factor of $P_{X-ray}^2/P_{OVI}^2$.
The maximum total X-ray and \OVI\ emitting pathlength through Loop I
under the two density assumptions would be either $\sim$ 2.5 pc or $\sim$
250 pc.

Assuming the distance to the IZ is $\sim$ 70 pc \citep{ea95}, and the
radius of the Loop I superbubble is $\sim$ 100 pc \citep{nish99}, we
can use the angular distance between the `Unshadowed' sightline and the
center of Loop I \citep{nish99} to geometrically determine the distance
to the center of Loop I and the pathlength of the `Unshadowed' sightline
through Loop I (under the poor assumption of spherical symmetry; see
Figure \ref{fig:schematic}).  We find that the pathlength along the
`Unshadowed' sightline is $\sim$15 pc, and the pathlength along the
`Shadowed' sightline is $\sim$ 20 pc. Although the exact numbers are
sensitive to the choice of input values and geometrical assumptions,
the main conclusion is not: since our sightlines do not lie near the
center of Loop I, our pathlength through Loop I must be substantially
less than its total diameter of $\sim$ 200 pc.  As a specific example, if
the distance to the IZ is larger than 70 parsecs, then the radius of the
(assumed) spherical Loop I must be larger; otherwise our `Unshadowed'
sightline lies outside of Loop I (unlikely given the measured \OVI\
intensity in this direction). The resulting inferred pathlength through
Loop I would then still be substantially less than the diameter of Loop I.
As a result, the density of the \OVI-emitting gas is more likely to be
$\sim$ 0.1 cm$^{-3}$ (pathlength $\sim$ 2.5 pc) than $\sim$ 0.01 cm$^{-3}$
(pathlength $\sim$ 250 pc).  We note that the thermal pressure of 30,000
cm$^{-3}$~K calculated under the assumption of $n \sim 0.1$ cm$^{-3}$
and $T \sim 300,000$~K compares well with the Loop I X-ray emitting gas
pressure of 46,000 cm$^{-3}$ K estimated by \citet{davelaar80} and the
midplane X-ray emitting gas pressure of 28,000 cm$^{-3}$~K determined
by \citet{snowden97}.  In addition, Figure 2 of \citet{breit06}, who
modelled the history and future of the Local and Loop I bubbles, also
suggests that the \OVI-emitting gas should be confined to the interior
interfaces of Loop I.

The intrinsic FWHM of the \OVI\ 1032 \AA\ emission line are reported in
Table \ref{table:OVI}.  Since an instrumental contribution of 25 km/s
corresponds to a convolving Gaussian with $\sigma_G$ = 0.037 \AA\, any
best-fit Gaussian narrower than this indicates that the LWRS aperture
is not filled: i.e. that the emitting region is non-uniform. This is
technically the case for our fit to the night-only spectrum along the
`Shadowed' sightline, but a filled aperture is within the 1-$\sigma$
error bars. Under the assumption that the entire intrinsic width is due
to thermal broadening, we calculated the temperature of the emitting
gas for each of our estimates of \OVI\ intensity.  For the night-only
`Shadowed' measurement, we assumed a width corresponding to the upper
end of the 1-$\sigma$ error range. This yields a temperature of $(1.9 \pm
1.4) \times 10^5$ K. The Day + Night `Shadowed' measurement corresponds
to a temperature of $(5.8 \pm 5.8) \times 10^5$ K. The night-only and
full-data `Unshadowed' measurements yielded temperatures of 3.9 $\pm$
0.15 million K and 2.5 $\pm$ 0.15 million K, respectively.  The `Shadowed'
measurements are roughly consistent with thermal broadening, while the
widths for the `Unshadowed' sightline are larger. This weakly suggests
that gas motions, possibly turbulent, are contributing to the width of
the \OVI\ emission in Loop I, or that the emission we see comes from
more than one distinct emitting region with different radial velocities.
As noted earlier, it is likely that the majority of the emission for even
the `Shadowed' sightline originates from beyond the IZ. The difference
in widths may indicate that the Loop I gas properties differ somewhat
between the sightlines.  Differences in best-fit radial velocities of the
`Shadowed' and `Unshadowed' emission also suggest inhomogeneous properties
for the emitting gas within Loop I.  It is important to note, however,
that non-uniform surface brightness within the 106 km/s wide aperture
profile can mimic a velocity shift, and the `Shadowed' sightline width
suggests this scenario. This could be due to either inhomogeneous emission
or patchy absorption by the IZ.

We note that our derived Loop I \OVI\ emission intensity is much less
than that measured by {\it SPEAR} towards the Vela Supernova Remnant
\citep{nish06}, as is expected since Vela is a much younger structure,
containing strong radiative shocks and filled with a large amount of gas
cooling through $\sim$ 3 $\times 10^5$ K.  {\it SPEAR}'s measurement
of \OVI\ towards the Orion-Eridanus superbubble is much more similar,
peaking at $\sim$ 7000 LU along the edge of the structure, and confined
to a region $\sim 5-10$ pc across \citep{kreg06}.  Thus the \OVI\ in
both Loop I and Orion-Eridanus appears to arise in interfaces at the
edge of the bubble, rather than from gas filling the interior.

\subsection{\CIII, low-ionization species, and the Local Bubble}

The observed \CIII\ intensities for the `Shadowed' and `Unshadowed'
sightlines are $I_{\rm shadowed}$ = 13300 $\pm$ 2200 \lu\ and $I_{\rm
unshadowed}$ = 15000 $\pm$ 5000 \lu.  These measurements are consistent
with the $\sim 10^4$ LU seen by {\it SPEAR} in this part of the sky
\citep{korpela06}, although the {\it SPEAR} measurement is an average
over 64 square degrees.  We assume that dust opacity at 977\AA\ is 13\%
higher than at 1032\AA\ \citep{sasseen02}, and use the same optical
depth analysis as for the \OVI.  For the \HI\ column densities from
the CHIANTI model of the previous section, we obtain \CIII\ intensities
of $12800\pm 3400$ \lu\ and  $3800^{+12300}_{-3800}$ \lu\ for emission
originating on the near and far sides of the IZ, respectively.  The large
error bars on the distant emission are primarily due to the opacity of
the intervening material.  The inferred nearby emission is substantially
more than the \CIII\ observed within the Local Bubble by \citet{shel03},
suggesting non-uniform emission, possibly as a result of \CIII\ production
at inhomogeneous interfaces.

Unlike \OVI, a significant proportion of the observed \CIII\ emission
arises on the near side of the IZ. Of the 15000 $\pm$ 5000 \lu\ observed
along the `Unshadowed' sightline, at least 9600 \lu\ (1-$\sigma$ lower
limit) arises within the Local Bubble. Such a situation could arise if
the interface between the Local Bubble and the interaction zone were
significantly cooler ($\lesssim 10^{5}K$) than the Loop I/interaction
zone interface, because hot gas within Loop I would further ionize
Carbon, reducing the amount of \CIII\ emission originating beyond the IZ.
This lends support to the idea that the Local Bubble is not filled with
overpressured hot ($10^6$ K) gas \citep{wel05}.  However, the large
error bars on $I_{beyond}$ prevent us from concluding that there is no
significant \CIII\ emission within Loop I.

Three other species, all present in the neutral ISM, have significant
measurements along both sightlines: \CII, \ion{Ar}{1}, and \ion{Fe}{2}. As
may be seen in Table \ref{table:otherlines}, in all cases the measurements
for the two sightlines are quite similar, agreeing within the errors. This
suggests that most of the emission arises in the near side of the IZ, as
for \CIII. Note that the \ion{Ar}{1} measurements may be contaminated by
nighttime airglow. The radial velocity for \ion{Ar}{1} along the `Unshadowed'
sightline is consistent with geocoronal emission at the 1-$\sigma$ level. 
Along the `Shadowed' sightline, the \ion{Ar}{1} radial velocity is inconsistent
with geocoronal emission at the 1-$\sigma$ (and possibly 2-$\sigma$) level, 
although it is difficult to be specific since that measurement combines
two observations with disparate geocentric to heliocentric corrections.
Apart from \ion{N}{2}, which may also be contaminated
by nighttime airglow, measurements and upper limits for all remaining
species in Table \ref{table:otherlines} are consistent between the two
sightlines, again suggesting that emission for these low-ionization
species arises largely in the near side of the IZ.

The radial velocities for these species are shown in Table
\ref{table:otherlines}. In general, the radial velocities along the
`Shadowed' sightline tend to be lower than along the `Unshadowed'
sightline, although the relatively large error bars make it difficult
to be specific. This could indicate that the emission along the
`Unshadowed' sightline arises at the boundary between the Local Bubble
and the surrounding neutral ISM, while the emission along the `Shadowed'
sightline arises at the interface between the Local Bubble and the
interaction zone. In this case, we would expect lower radial velocities
along the `Shadowed' sightline, because the expansion of Loop I could
push the IZ towards us.

\section{Conclusions}

We have performed a shadowing measurement using absorption of \OVI\
emission along a line of sight that intersects the purported interaction
zone between the Loop I superbubble and the Local Bubble.  The results
are consistent with moderately bright \OVI\ emission ($\sim 10^4$ \lu)
interior to Loop I, likely arising near the walls of the bubble, with
absorbing dust in the interaction region equivalent to an \HI\ column
density of $\sim 4\times10^{20} \percmsqr$. Less than $\sim$ 800 \lu\
of the observed emission arises from the interface between the Local
Bubble and the interaction zone.

Unlike \OVI, emission from \CIII\ and other low-ionization species
along the same sightlines shows a very significant local component. The
disparity in the origin of \OVI\ and \CIII\ on these sightlines indicates
that the conditions in the interface between the interaction region
and the Local Bubble are significantly different than in the hot/cold
interfaces within Loop I.  These differences are likely due to differences
in the temperature and pressure of Local Bubble gas relative to conditions
within Loop I.






\acknowledgments

We are grateful to W. Corradi \& W. Reis for providing a detailed
reddening map of the region near our observations, and K. Nishikida
for discussions about Loop I.  This paper used observations obtained by
the NASA-CNES-CSA {\it Far Ultraviolet Spectroscopic Explorer} (\FUSE)
mission operated by the Johns Hopkins University, supported by NASA
contract NAS5-32985.  W.V.D. Dixon was extremely helpful with questions
about the \FUSE\ pipeline.  Funding for this project was provided by NASA
through NASA grant NNG05GB62G. We acknowledge the use of NASA's {\it SkyView}
facility (http://skyview.gsfc.nasa.gov) located at NASA Goddard Space 
Flight Center. This research has made use of data obtained from the High 
Energy Astrophysics Science Archive Research Center (HEASARC), provided by 
NASA's Goddard Space Flight Center.
{\it CHIANTI} is a collaborative project involving the NRL (USA), 
RAL (UK), and the following Univerisities: College London (UK), Cambridge (UK), 
George Mason (USA), and Florence (Italy). 




Facilities: \facility{\FUSE}.

\begin{deluxetable}{lccccccc}
\tablewidth{0pt}
\tabletypesize{\small}
\tablecaption{1032 \AA\ \OVI\ Emission \label{table:OVI}}
\tablehead{
\colhead{Sightline}     &\colhead{Data} & 
\colhead{\OVI\ intensity} &
\colhead{$\sigma_G$\tablenotemark{c}} 	& 	\colhead{FWHM\tablenotemark{c}} &
\colhead{V$_{\rm LSR}$} \\
\colhead{}     &\colhead{} & 
\colhead{(LU\tablenotemark{b})}  &
\colhead{(\AA)} 	& 	\colhead{(\kms)} &
\colhead{(\kms)} 
}
\startdata
Shadowed\tablenotemark{a} & Night & 2750 $\pm$ 550 & 0.02$^{+.03}_{-0.02}$ & $\leq$ 21\tablenotemark{d} & $-$29 $\pm$ 6 \\
... & Day + Night & 2640 $\pm$ 600 & 0.07 $\pm$ 0.06 & 41 $\pm$ 41 & $-$29 $\pm$ 10 \\
Unshadowed\tablenotemark{a} &Night & 10800 $\pm$ 1200 & 0.16 $\pm$ 0.03& 107 $\pm$ 21& 15  $\pm$ 8 \\
... &Day + Night & 9580 $\pm$ 1000 & 0.13 $\pm$ 0.03 & 85 $\pm$ 21 & 16  $\pm$ 6 \\
\enddata
\tablenotetext{a}{Shadowed: ($l,b$) = (278.23, +8.02), Unshadowed: ($l,b$) =
(276.26, +10.69)}
\tablenotetext{b}{\ LU = \lu}
\tablenotetext{c}{FWHM have been corrected for instrumental contribution, while
$\sigma_G$ have not. Gas at $T = 3 \times 10^5$ K has a thermal width corresponding to a FWHM of 29 km/s.}
\tablenotetext{d}{The best-fit convolving Gaussian has a FWHM less than the
instrumental contribution, suggesting the aperture is not filled, although a
filled aperture is within the error bars.}
\end{deluxetable}

\begin{deluxetable}{llccccccc}
\rotate
\center
\tablewidth{0pt}
\tabletypesize{\tiny}
\tablecaption{Other Emission Lines and Upper Limits \label{table:otherlines}}
\tablehead{
\multicolumn{2}{l}{}  & \multicolumn{1}{c}{} & 
\multicolumn{3}{c}{SHADOWED} & \multicolumn{3}{c}{UNSHADOWED} \\
\colhead{Species}  & \colhead{Wavelength} & \colhead{Detector} &
\colhead{Intensity}  & \colhead{$\sigma_G$\tablenotemark{c}} & \colhead{$V_{LSR}$} & 
\colhead{Intensity}  & \colhead{$\sigma_G$\tablenotemark{c}} & \colhead{$V_{LSR}$} \\
\colhead{}  & \colhead{(\AA)} & \colhead{Segment} &
\colhead{(LU\tablenotemark{b})}  & \colhead{(\AA)}  & \colhead{(km s$^{-1}$)} & 
\colhead{(LU)}  & \colhead{(\AA)} & \colhead{(km s$^{-1}$)} }
\startdata
C I\tablenotemark{a} & 945.58 & SiC 2a & 2900 $\pm$ 1500 & ... & ... & $<$ 8700 & ... & ... \\
Mg II  & 946.73 & SiC 2a & 2500 $\pm$ 1000 & ... & $-26 \pm 13$ & $<$8700 & ... & ... \\
N I\tablenotemark{a} & 953.97 & SiC 2a & $<$4400 & ... & ... & $<$8800 & ... & ... \\
C III & 977.02 & SiC 2a & 13300 $\pm$ 2200  & 0.14 $\pm$ 0.04 & $-10 \pm 10$ & 15000$\pm$ 5000 & 0.29 $\pm$ 0.14 & 45 $\pm$ 35 \\
N III & 991.57 & SiC 2a & $<$5300 & ... & ... & $<$10600 & ... & ... \\
S III & 1015.55 & LiF 1a & $<$1800 & ... & ... & $<$3200 & ... & ... \\
C II & 1037.02 & LiF 1a & 5100 $\pm$ 1100  & 0.28 $\pm$ 0.12 & 8 $\pm$ 16 & 6700$\pm$ 1300 & 0.21 $\pm$ 0.06 & 13 $\pm$ 16 \\
Ar I\tablenotemark{a} & 1048.22 & LiF 1a & 5000 $\pm$ 1100 & 0.24 $\pm$ 0.07 & $-61 \pm 19$ & 5200 $\pm$ 1200 & 0.17 $\pm$ 0.06 & $-14 \pm 16$ \\
S IV & 1062.66 & LiF 1a & 1200 $\pm$ 800 & ... & 94 $\pm$ 45 & $<$3300 & ... & ... \\
S IV & 1072.99 & LiF 1a & $<$2500 & ... & ... & $<$4000 & ... & ... \\
N II\tablenotemark{a} & 1083.99\tablenotemark{d} & SiC 1a & $<$12000 & ... & ... & 20000 $\pm$ 6000 & ... & ... \\
C I\tablenotemark{a} & 1122.26 & LiF 2a & $<$2300 & ... & ... & $<$3400 & ... & ... \\
Fe III & 1122.52 & LiF 2a & $<$2300 & ... & ... & 3100 $\pm$ 1400 & ... & $-17 \pm 35$ \\
Fe II & 1144.94 & LiF 2a & 2900 $\pm$ 800 & 0.05 $\pm$ 0.07 & $-26 \pm 13$ & 2700 $\pm$ 1300 & 0.1 $\pm$ 0.1 & 23 $\pm$ 22 \\
\enddata
\tablenotetext{a}{These lines may contain a contribution due to nighttime airglow}
\tablenotetext{b}{\ LU = \lu}
\tablenotetext{c}{The width of the best-fit convolving Gaussian is not included
for lines fit only because they exceeded the 95\% confidence limit, as these
were poorly determined.}
\tablenotetext{d}{Because it was not possible accurately determine the
wavelength zero-point of the SiC 1a spectrum, we are unable to determine with
certainty which of the NII multiplet lines we observed.}
\end{deluxetable}




\end{document}